\documentclass{article}

\usepackage{epsfig}
\usepackage{subfigure}
\usepackage{calc}
\usepackage{graphicx}

\usepackage{amsfonts}
\usepackage{amsmath}
\usepackage{amssymb}
\usepackage{latexsym}

\begin{document}

\title{Simple Cellular Automata-Based Linear Models for the Shrinking Generator}
\date{}
\author{Amparo F\'{u}ster-Sabater$^{(1)}$ and Dolores de la Gu\'{\i}a-Mart\'{\i}nez$^{(2)}$\\
{\small (1) Instituto de F\'{\i}sica Aplicada, C.S.I.C.}\\
{\small Serrano 144, 28006 Madrid, Spain} \\
{\small amparo@iec.csic.es}\\
{\small (2) Centro T\'{e}cnico de Inform\'{a}tica, C.S.I.C.} \\
{\small Pinar 19, 28006 Madrid, Spain} \\
{\small lola@cti.csic.es }}

\maketitle

\begin{abstract}
Structural properties of two well-known families of keystream
generators, Shrinking Generators and Cellular Automata, have been
analyzed. Emphasis is on the equivalence of the binary sequences
obtained from both kinds of generators. In fact, Shrinking
Generators (SG) can be identified with a subset of linear
Cellular Automata (mainly rule 90, rule 150 or a hybrid
combination of both rules). The linearity of these cellular models
can be advantageously used in the cryptanalysis of those keystream
generators.
\end{abstract}

\section{Introduction}
\footnotetext{This work has been supported by Ministerio de
Ciencia y Tecnolog\'{\i}a (Spain) under grant TIC2001-0586. \\
Proceedings 2003 IEEE Information Theory Workshop, pp. 143-146. La
Sorbonne, Paris,  31 March - 4 April, 2003.} Cellular Automata
(CA) are discrete dynamic systems characterized by a simple
structure but a complex behavior \cite{Mar, Nan, Da}. This
configuration makes them very attractive to be used in the
generation of pseudorandom sequences. In this sense, CA are
studied in order to obtain a characterization of the rules
(mapping to the next state) producing sequences with maximal
length, balancedness and good distribution of 1's and 0's. From a
cryptographic point of view, it is fundamental to analyze some
additional characteristics of these generators, such as linear
complexity or correlation-immunity. The results of this study
point toward the equivalence between the sequences generated by CA
and those obtained from Linear Feedback Shift Registers-based
models \cite{Go}.

In this paper, CA hybrid configurations constructed from
combinations of rules 90 and 150 are considered. In fact, a linear
model that describes the behavior of a kind of pseudorandom
sequence generator, the so-called shrinking generator (SG), has
been derived. In this way, the sequences generated by SG can be
studied in terms of CA. Thus, all the theoretical background on CA
found in the literature can be applied to the analysis and/or
cryptanalysis of shrinking generators.

\section{General description of the basic structures}

The two basic structures are introduced:

\subsection{The Shrinking Generator}

It is a very simple generator with good cryptographic properties
\cite{Co}. This generator is composed by two LFSRs: a control
register, called $R_{1}$, that decimates the sequence produced by
the other register, called $R_{2}$. The sequence produced by the
LFSR $R_{1}$, that is $\{a_{i}\}$, controls the bits of the
sequence produced by $R_{2}$, that is $\{b_{i}\}$, which are
included in the output sequence $\{c_{j}\}$ (the shrunken
sequence) , according to the following rule:

\begin{enumerate}
\item  If $a_{i}=1\Longrightarrow c_{j}=b_{i}$

\item  If $a_{i}=0\Longrightarrow b_{i}$ is discarded.
\end{enumerate}

\textit{Example 1: }Consider the following LFSRs:

\begin{enumerate}
\item  $R_{1}$ with length $L_{1}=3$, feedback polynomial $1+D+D^{3}$ and
initial state $(1,0,0)$. The sequence obtained is $%
\{0,0,1,1,1,0,1\}$ with period $7$.

\item  $R_{2}$ with length $L_{2}=4$, feedback polynomial $1+D^{3}+D^{4}$ and
initial state $(1,0,0,0)$. The sequence obtained is $%
\{0,0,0,1,0,0,1,1,0,1,0,1,1,1,1\}$ with period $15$.
\end{enumerate}

The output sequence $\{c_{j}\}$ will be determined by:

\begin{itemize}
\item  $\{a_{i}\}$ $\rightarrow $ $0\;0\;1\;1\;1\;0\;1\;0\;0\;1\;1\;1\;0\;1%
\;0\;.....$

\item  $\{b_{i}\}$ $\rightarrow $ \underline{$0$}$\;\underline{0}\;0\;1\;0\;%
\underline{0}\;1\;\underline{1}\;\underline{0}\;1\;0\;1\;\underline{1}\;1\;\underline{1}%
\;.....$

\item  $\{c_{j}\}$ $\rightarrow $ $0\;1\;0\;1\;1\;0\;1\;1\;\;.....$
\end{itemize}

The underlined bits \underline{0} or \underline{1} in $\{b_{i}\}$
are discarded.

According to \cite{Co}, the period of the shrunken sequence is
\begin{equation}
T=(2^{L_{2}}-1)2^{(L_{1}-1)}
\end{equation}
and its linear complexity, notated $LC$, satisfies the following
inequality
\begin{equation}
L_{2} \thinspace 2^{(L_{1}-2)}<LC\leq L_{2} \thinspace
2^{(L_{1}-1)}.
\end{equation}
A simple calculation allows one to compute the number of $1$'s in
the shrunken sequence. Such a number is
\begin{equation} No. 1's = 2^{(L_{2}-1)}2^{(L_{1}-1)}.
\end{equation}
%\bigskip
Thus, the shrunken sequence is a quasi-balanced sequence. Since
simplicity is one of its most remarkable characteristics, this
scheme is suitable for practical implementation of efficient
stream cipher cryptosystems.

\subsection{Cellular Automata}

An one-dimensional cellular automaton can be described as a
\textit{n}-cell register, whose binary stages are updated at the
same time depending on a \textit{k}-variable function \cite{Da}
also called \textit{rule}. If $k=2r+1$ input variables are
considered, then there is a total of $2^{k}$ different
neighborhood configurations. Therefore, for a binary cellular
automaton there can be up to $2^{2^{k}}$ different mappings to the
next
state. Such mappings are the different rules $\Phi $.  In fact, the next state $%
x_{i}^{t+1}$ of the cell $x_{i}^{t}$ depends on the current state
of $k$ neighbor cells
\begin{equation}
x_{i}^{t+1}=\Phi (x_{i-r}^{t},\ldots ,x_{i}^{t},\ldots
,x_{i+r}^{t})
\end{equation}
If these functions are composed exclusively by XOR and/or XNOR
operations, then CA are said to be \textit{additive}. In this case, the next state $%
(x_{1}^{t+1},\ldots ,x_{n}^{t+1})$ can be computed from the current state $%
(x_{1}^{t},\ldots ,x_{n}^{t})$ such as follows:
\begin{equation}
(x_{1}^{t+1},...,x_{n}^{t+1})=(x_{1}^{t},...,x_{n}^{t}).A+C
\end{equation}

where $A$ is an $n\thinspace x \thinspace n$ matrix with binary
coefficients and $C$ is the complementary vector.

In CA, either all cells evolve under the same rule
(\textit{uniform case}) or they follow different rules
(\textit{hybrid case}). At the ends of the array, two different
boundary conditions are possible: \textit{null automata} whether
cells with permanent null content are supposed adjacent to the
extreme cells or \textit{periodic automata} whether extreme cells
are supposed adjacent.

In this paper, all automata considered will be null hybrid CA with
rules 90 y 150. For $k=3$, these rules are described such as
follows :

\begin{itemize}
\item  rule 90 \thinspace $\rightarrow
x_{i}^{t+1}=x_{i-1}^{t}+x_{i+1}^{t}$\\
$
\begin{array}{cccccccc}
111 & 110 & 101 & 100 & 011 & 010 & 001 & 000 \\
0 & 1 & 0 & 1 & 1 & 0 & 1 & 0
\end{array}
$

$01011010$ (binary) = $90$ (decimal).

\item  rule 150 $\rightarrow
x_{i}^{t+1}=x_{i-1}^{t}+x_{i}^{t}+x_{i+1}^{t}$\\
$
\begin{array}{cccccccc}
111 & 110 & 101 & 100 & 011 & 010 & 001 & 000 \\
1 & 0 & 0 & 1 & 0 & 1 & 1 & 0
\end{array}
$

$10010110$ (binary) = $150$ (decimal).
\end{itemize}

The main idea of this work is to write a given SG in terms of a
hybrid cellular automaton, where at least one of its output
sequences equals the SG output sequence.

\section{A shrinking generator linear model in terms of cellular automata}

In this section, an algorithm to determine the one-dimensional
linear hybrid CA corresponding to a particular shrinking generator
is presented. Such an algorithm is based on the following facts:

\begin{description}
\item  \textbf{Fact 1:} The characteristic polynomial of the
shrunken sequence \cite{Co} is of the form
\begin{equation}
P(D)^{N}
\end{equation}
where $P(D)$ is a $L_{2}$-degree primitive polynomial and $N$
satisfies the inequality $2^{(L_{1}-2)}<N\leq 2^{(L_{1}-1)}$.

\item  \textbf{Fact 2:} $P(D)$ depends exclusively on the characteristic
polynomial of the register $R_{2}$ and on the length $L_{1}$ of
the register $R_{1}$. Moreover, $P(D)$ is the characteristic
polynomial of cyclotomic \textit{coset} $2^{L_{1}}-1$, see
\cite{Go}. This result can be proved in the same way as the lower
bound on the $LC$ is derived in reference \cite{Co}.

\item  \textbf{Fact 3:} Rule $90$ at the end of the array in a null automaton
is equivalent to two consecutive rules $150$ with identical
sequences. Reciprocally, rule $150$ at the end of the array in a
null automaton is equivalent to two consecutive rules $90$ with
identical sequences.
\end{description}

According to the previous facts, the following algorithm is
introduced:

\textit{Input:} Two LFSR's $R_{1}$ and $R_{2}$ with their
corresponding lengths, $L_{1}$ and $L_{2}$, and the characteristic
polynomials $P_{2}(D)$ of the register $R_{2}$.

\begin{description}
\item  \textit{Step 1:} From $L_{1}$ and $P_{2}(D)$, compute the polynomial $P(D)$.
In fact, $P(D)$ is the characteristic polynomial of the cyclotomic
\textit{coset E}, where
%\[
$E =2^{0}+2^{1}+\ldots +2^{L_{1}-1}$.
%\]
Thus,
%\[
$P(D) =(D+\alpha^{E})(D+\alpha^{2E})\ldots
(D+\alpha^{2^{L_{1}-1}E})$
%\]
$\alpha$ being a primitive root in $GF(2^{L_{2}})$.

\item  \textit{Step 2:} From $P(D)$, apply the Cattell and Muzio
synthesis algorithm \cite{Ca} to determine the two linear hybrid
CA whose characteristic polynomial is $P(D)$. Such CA are written
as binary strings with the following codification: $0$ = rule $90$
and $1$ = rule $150$.

\item  \textit{Step 3:} For each one of the previous binary strings representing the
CA, we proceed:
\begin{description}
  \item \textit{3.1} Complement its least significant bit. The resulting binary string is notated $S$.

  \item \textit{3.2} Compute the mirror image of $S$, notated $S^{*}$, and
  concatenate both strings
  %\[
  $S_{c} = S*S^{*}$.
  %\]
  \item \textit{3.3} Apply steps $3.1$ and $3.2$ to $S_{c}$ recursively $L_{1}-1$ times.
\end{description}

\end{description}

\textit{Output: } Two binary strings codifying the CA
corresponding to the given SG.

Remark that the characteristic polynomial of the register $R_{1}$
is not needed. Due to the particular form of the shrunken sequence
characteristic polynomial, it can be noticed that the computation
of the CA is proportional to $L_{1}$ instead of $2^{L_{1}}$.
Consequently, the algorithm can be applied to SG in a range of
cryptographic interest (e.g. $L_{1}, L_{2}\approx 64$). In order
to clarify the previous steps a simple numerical example is
presented.

\bigskip
\textit{Example 2: }Consider the following LFSRs: $R_{1}$ with
length $L_{1}=2$ and $R_{2}$ with length $L_{2}=5$ and
characteristic polynomial $P_{2}(D)=1+D+D^{3}+D^{4}+D^{5}$.

\begin{description}
\item  \textit{Step 1:} $P(D)$ is the characteristic polynomial of the cyclotomic
\textit{coset 3}. Thus,
%\[
$P(D) =1+D^{2}+D^{5}$.
%\]

\item  \textit{Step 2:} From $P(D)$ and applying the Cattell and Muzio
synthesis algorithm, two linear hybrid CA whose characteristic
polynomial is $P(D)$ can be determined. Such CA are written as:
\begin{center}
$
\begin{array}{ccccc}
0 & 1 & 1 & 1 & 1 \\
1 & 1 & 1 & 1 & 0
\end{array}
$
\end{center}

\item  \textit{Step 3:} The two binary strings of length $L = 10$ representing the
required CA are:
\begin{center}
$
\begin{tabular}{llllllllll}
0 & 1 & 1 & 1 & 0 & 0 & 1 & 1 & 1 & 0 \\
1 & 1 & 1 & 1 & 1 & 1 & 1 & 1 & 1 & 1
\end{tabular}
$
\end{center}

with the corresponding codification above mentioned. The procedure
has been carried out once as $L_{1}-1 = 1$.
\end{description}

From $L = 10$ known bits of the shrunken sequence $\{c_{j}\}$, the
whole sequence, whose period $T=62$, can be easily reconstructed.
In fact, let $\{c_{j}\}$ be of the form
\begin{center}
$\{c_{j}\} = \{0\;1\;0\;1\;1\;0\;1\;0\;0\;1\;...\},$
\end{center}
then the initial state of the cellular automaton can be computed
from right to left (or viceversa), according to the corresponding
rules $90$ and $150$. Tables 1 depicts the computation of the
initial state for the first automaton. The shrunken sequence is
placed at the most right column.
\begin{center}
 $
\begin{array}{cccccccccc}
\textbf{90} &\textbf{150} &\textbf{150} &\textbf{150} &\textbf{90} &\textbf{90} &\textbf{150} &\textbf{150} &\textbf{150} &\textbf{90} \\
0 & 0 & 0 & 1 & 1 & 1 & 0 & 1 & 1 & 0 \\
& 0 & 1 & 0 & 0 & 1 & 0 & 0 & 0 & 1 \\
&  & 1 & 1 & 1 & 0 & 1 & 0 & 1 & 0 \\
&  &  & 1 & 1 & 0 & 1 & 0 & 1 & 1 \\
&  &  &  & 1 & 0 & 1 & 0 & 0 & 1 \\
&  &  &  &  & 0 & 1 & 1 & 1 & 0 \\
&  &  &  &  &  & 0 & 1 & 0 & 1 \\
&  &  &  &  &  &  & 1 & 0 & 0 \\
&  &  &  &  &  &  &  & 1 & 0 \\
&  &  &  &  &  &  &  &  & 1
\end{array}
$
\end{center}
\vspace*{-0.6cm}
\begin{center}
{\em Table. 1 - The shrunken sequence is at the most right column}
\end{center}

Once the corresponding initial states are known, then the cellular
automata will produce their corresponding output sequences and the
shrunken sequence can be univocally determined.

In fact, CA computed by the previous algorithm will generate all
the possible sequences $\{x_{i}\}$ that are solutions of the
difference equation
\begin{equation}
[P(E)]^{2^{L_1 -1}}\{x_{i}\}=0
\end{equation}

$E$ being the shifting operator on $x_{i}$ (i.e.
$Ex_{i}=x_{i+1}$). The shrunken sequence $\{c_{j}\}$ is just a
particular solution of the previous equation. The different
sequences $\{x_{i}\}$ are distributed into the different state
cycles of each automaton. Once a specific sequence is fixed in a
particular cell, e. g. the shrunken sequence at the most right
cell in the previous example, the location of the other sequences
is univocally determined. In addition, every particular solution
$\{x_{i}\}$ can be generated by every automaton cell depending on
the state cycle considered. In terms of LFSR-based generators, the
solution sequences $\{x_{i}\}$ correspond to sequences generated
by different combinations of LFSRs: clock-controlled shrinking
generators \cite{Kan}, shrinking generators with distinct rules of
decimation, irregular clocking of the register $R_{2}$ based on
particular stages of the register $R_{1}$ etc. All these
generators are included in a simple automaton.

\section{Applications of the CA-based model to the cryptanalysis
of the shrinking generator}

Since a linear model describing the behavior of the SG has been
derived, the cryptanalysis of such a generator can be considered
from different points of view:

\begin{itemize}
  \item Crytanalysis based on the SG linear complexity: attacking the SG through
  its linear complexity requires the knowledge of $LC$ bits of the shrunken sequence,
  $LC$ being its linear complexity, or equivalently, the length of the cellular automaton.
  Remark that this is just half the sequence required by the
  Berlekamp-Massey algorithm \cite{Ma} to reconstruct the original sequence.

  \item Crytanalysis based on the Linear Consistency Test (LCT): The linear consistency test \cite{Ze} is a
  general divide-and-conquer cryptanalytic attack that can be applied to the SG
  on the basis of the linear models
  provided by the cellular automata. This attack would require the exhaustive
  search through all possible initial states of the LFSR $R_{2}$.

  \item A new attack that exploits the weaknesses inherent to the CA-based
  linear model can be also considered. Such an attack will be
  specified in next section.

\end{itemize}

\section{Phaseshift analysis of CA sequences}

If the Bardell's algorithm to phaseshift analysis of CA \cite{Bar} is
applied, then it is possible to calculate the relation among the
sequences obtained from CA. In fact, in \cite{Bar} it was shown that the characteristic equation determines the
recursion relationship among the bits in the output sequences of a
hybrid 90/150 CA. A shift operator was used in conjunction with a
table of discrete logarithms to determine the phaseshift
analytically.

Although the characteristic equation in \cite{Bar} is a primitive
polynomial $P(D)$, it can be proved that the algorithm is valid for
$P(D)^n$ too.

\bigskip
\textit{Example 3: }Let us consider a CA with the following characteristics:
\begin{itemize}
  \item The automaton length $L=10$
  \item Rule distribution: $0011001100$
  \item $P(D)=(1+D+D^{2}+D^{4}+D^{5})^{2}$.
  \end{itemize}

  Let $S$ be a shift operator defined in $GF(2)$ which operates on
  $X_{i}$, the state of the \textit{i}-th cell , such as follows:
\begin{equation}
X_{i}(t+1)= SX_{i}(t)
\end{equation}
we can write
\begin{equation}
X_{1}(t+1)= X_{2}(t)
\end{equation}
as
\begin{equation}
SX_{1}(t)= X_{2}(t)
\end{equation}
or simply
\begin{equation}
SX_{1}= X_{2}.
\end{equation}

The difference equation system is as follows:
\begin{eqnarray*}
SX_{1} &=&X_{2} \\
SX_{2} &=&X_{1}+X_{3} \\
SX_{3} &=&X_{2}+X_{3}+X_{4} \\
SX_{4} &=&X_{3}+X_{4}+X_{5} \\
SX_{5} &=&X_{4}+X_{6} \\
SX_{6} &=&X_{5}+X_{7} \\
SX_{7} &=&X_{6}+X_{7}+X_{8} \\
SX_{8} &=&X_{7}+X_{8}+X_{9} \\
SX_{9} &=&X_{8}+X_{10} \\
SX_{10} &=&X_{9}
\end{eqnarray*}

Expressing each $X_{i}$ as a function of $X_{10}$, we obtain the
following system:
\begin{eqnarray*}
X_{1} &=& (S^9+S^4+S^3+S^2+S+1)X_{10} \\
X_{2} &=& (S^8+S^6+S^5+S^4+S^3+S+1)X_{10} \\
X_{3} &=& (S^7+S^6+S^5+S^3+1)X_{10} \\
X_{4} &=& (S^6)X_{10} \\
X_{5} &=& (S^5+S^3+1)X_{10} \\
X_{6} &=& (S^4+S)X_{10} \\
X_{7} &=& (S^3+S^2+1)X_{10} \\
X_{8} &=& (S^2+1)X_{10} \\
X_{9} &=& (S)X_{10}
\end{eqnarray*}
Now taking logarithms in both sides,
\begin{eqnarray*}
log(X_{1})&=&log(S^9+S^4+S^3+S^2+S+1)+log(X_{10}) \\
log (X_{2})&=&log(S^8+S^6+S^5+S^4+S^3+S+1)+ \\
&+&log(X_{10}) \\
log(X_{3})&=&log(S^7+S^6+S^5+S^3+1)+log(X_{10}) \\
log (X_{4})&=&log(S^6)+log(X_{10}) \\
log(X_{5})&=&log(S^5+S^3+1)+log(X_{10}) \\
log(X_{6})&=&log(S^4+S)+log(X_{10}) \\
log (X_{7})&=&log(S^3+S^2+1)+log(X_{10}) \\
log(X_{8})&=&log(S^2+1)+log(X_{10}) \\
log (X_{9})&=&log(S)+log(X_{10})
\end{eqnarray*}

On the other hand, we have:
\begin{equation}
D^{26}\thinspace mod \thinspace P(D) = D^2+1.
\end{equation}
Next we define,
\begin{equation}
log (D) \equiv 1
\end{equation}

According to the algorithm proposed by Bardell, we can identify
the following equations:
\begin{eqnarray*}
log (X_{9}) - log (X_{10}) &=& 1 \\
log (X_{8}) - log (X_{10}) &=& 26 \\
log (X_{4}) - log (X_{10}) &=& 6
\end{eqnarray*}
and,
\begin{eqnarray*}
log (X_{2}) - log (X_{1}) &=& 1 \\
log (X_{3}) - log (X_{1}) &=& 26 \\
log (X_{7}) - log (X_{1}) &=& 6
\end{eqnarray*}

According to the previous results, the same binary sequence is generated in cells
 1, 2, 3 and 7 as well as the same sequence is produced in cells 10, 9, 8 and 4. The phaseshifts of the outputs
2, 3 and 7 relative to cell 1 are 1, 26 and 6 respectively.
Similar values are obtained in the other group of cells, that is
cells 4, 8 and 9 relative to cell 10. The rest of cells generate
different sequences.

Studying the distance
  among the shifted sequences and concatenating them, the original
  sequence can be reconstructed. Nevertheless, the shifts among the
  different shrunken sequences depend on the particular structure
  of the automaton considered. In fact, once the automaton is
  known the Bardell's algorithm has to be applied.

\bigskip
\section{Conclusions}

In this paper, the relationship between LFSR-based structures and
cellular automata have been stressed. More precisely, a particular
family of LFSR-based keystream generators, the so-called Shrinking
Generators, has been analyzed and identified with a subset of
linear cellular automata. In fact, a linear model describing the
behavior of the SG has been derived.

The algorithm to convert the SG into a CA-based linear model is
very simple and can be applied to shrinking generators in a range
of practical interest. The key idea of this algorithm is that the
number of steps to be carried out is proportional to $L_{1}$
instead of $2^{L_{1}}$.

Once the linear equivalent model has been developed, the linearity
of this cellular model can be advantageously used in the analysis
and/or cryptanalysis of the SG. Besides the traditional
cryptanalitic attacks (e.g. the linear complexity attack that here
requires half the sequence needed by the Berlekamp-Massey
algorithm and the LCT attack), an outline of a new attack that
exploits the weaknesses inherent to these CA has been introduced
too.

The proposed linear model is believed to be a very useful tool to
analyze the strength of the sequence produced by a SG as a
keystream generator in stream ciphers procedures.

\end{document}